# Room-temperature magnetoelectric coupling in strontium titanate

Zhen Yang[1,2][†], Cheng Ma[1,2][†], Mengqin Wang[1,2][†], Yuzhou He[1,2], Yuansha Chen[1,2], Fengxia Hu[1,2], Ye Yuan[1,2], Shuai Xu[1,2], Chen Ge[1,2], Er-jia Guo[1,2], Can Wang[1,2], Xiulai Xu[1,3], Guozhen Yang[1,2], Qinghua Zhang[1,2]*, and Kui-juan Jin[1,2]*

*1 Institute of Physics, Chinese Academy of Sciences, Beijing 100190, China*

*2 University of Chinese Academy of Sciences, Beijing 100049, China*

*3 State Key Laboratory for Mesoscopic Physics and Frontiers Science Center for Nano-optoelectronics, School of Physics, Peking University, Beijing 100871, China*

[*]Corresponding author: zqh@iphy.ac.cn (Qinghua Zhang); kjjin@iphy.ac.cn (Kui-juan Jin)

[†]Zhen Yang, Cheng Ma, and Mengqin Wang contributed equally to this work.

**Abstract**

Magnetoelectric (ME) multiferroics enable efficient interconversion of electrical and magnetic signals, offering pathways toward substantial reduction of power consumption in next-generation computing and information technologies. However, despite decades of research, the persistence of ME coupling at room-temperature, an indispensable feature for practical applications, remains exceedingly rare in single-phase materials, due to symmetry constraints imposed by magnetic point groups and competing electronic requirements for ferroelectricity and magnetism. Here we report the coexistence of ferroelectricity and ferromagnetism at 780 K and a strong ME coupling with a converse ME coupling coefficient up to 498 ps/m at room-temperature in strontium titanate by dedicated vacancy engineering. The asymmetric distribution of oxygen vacancies surrounding titanium vacancies enables atomic displacement of titanium and charge injection, providing joint origin for ferroelectricity and ferromagnetism, as confirmed by atomic-scale structural analysis and first-principles calculations. The mechanism of ME coupling is offered as the hopping of oxygen vacancies under electric fields, which leads to the rearrangement of electronic configuration. Our work opens an avenue for designing multiferroics, overcoming the long-standing scarcity of room-temperature single-phase multiferroics.

The rapid expansion of the computing and information industry and the collapse of Moore's law and Dennard scaling are dramatically converging in this decade, making the development of next-generation electronics an urgent necessity[1-4]. Magnetoelectric (ME) multiferroics, which exhibit both magnetic and electric orders, enable the manipulation of magnetic states via electric fields and vice versa, serving as a crucial interface for spintronics and promising substantial reductions in power consumption of writing and reading operations[5-14]. However, the simultaneous breaking of spatial and temporal symmetries, along with the inherent incompatibility between ferroelectricity and magnetism due to electronic structure conflicts, has led to a scarcity of room-temperature single-phase ME materials despite decades of research[7-10]. Although a number of ME composites incorporating ferroelectric and magnetic materials have been fabricated[15-18], single-phase ME multiferroics not only exhibit superiority in terms of precise modulation and industrial manufacturing but also provide deeper insights into the mechanism of ME coupling owing to their relatively simple crystal structures[10,11,19]. Therefore, the design and fabrication of room-temperature single-phase ME materials remain a significant challenge and a critical research topic.

Here we report the coexistence of ferroelectricity and ferromagnetism in strontium titanate ($SrTiO_3$, STO) by dedicated vacancy engineering, with ferroelectric Curie temperature ($T_C^{FE}$) higher than 780 K and ferromagnetic Curie temperature ($T_C^{FM}$) of 830 K. Above all, converse ME coupling (magnetization controlled by electric field) at room-temperature was achieved, with a converse ME coupling coefficient $\alpha_{CME}$ up to 498 ps/m, which is superior to the previous reported single-phase ME materials[7,20,21]. The microscopic mechanism involving the distribution of O vacancies surrounding Ti vacancies is revealed by a combination of optical second harmonic generation (SHG) measurements, electron ptychography and high-angle annular dark-field (HAADF) images, and first-principles calculations.

The STO films with thicknesses of 200 nm (Supplementary Fig. 1) were epitaxially grown on the (001)-$Sr_{1-x}Nb_xTiO_3$ ($x$ = 0.007) substrates via pulsed laser deposition under various oxygen pressure ($P_{oxygen}$) (details in methods), with high-quality epitaxial growth confirmed by X-ray diffraction (XRD) $\theta$-$2\theta$ symmetric scans (Fig. 1a) and reciprocal space mappings (RSM) (Supplementary Fig.1). The out-of-plane lattice constants $c$ were extracted as shown in the inset of Fig. 1a. Compared with bulk STO, these films exhibit larger $c$ due to the constraint of in-plane lattice constants and the lattice expansion caused by the presence of O vacancies, which is confirmed by the X-ray photoelectron spectroscopy (XPS) fitting results of O 1$s$ characteristic spectra (Fig. 1b and Supplementary Fig. 2). Meanwhile, the resultant $Ti^{3+}$ states are also captured by those of Ti 2$p$ characteristic spectra (Fig. 1c and Supplementary Fig. 3). To determine the elemental compositions of these films, Rutherford backscattering spectroscopy (RBS) measurements were conducted, demonstrating the deficiency of both O and Ti elements as shown in Fig. 1d (Supplementary Fig. 4 and Table 1)[22,23]. By normalizing the composition of Sr to 1, the chemical formula $SrTi_{1-\beta}O_{3-\gamma}$ were determined from the stoichiometric ratio, with $\gamma$ and $\beta$ representing the contents of O and Ti vacancies, respectively. As shown in Fig. 1e, the contents of O and Ti vacancies are synchronously modulated by $P_{oxygen}$, while their ratio ($\beta/\gamma$) remains unchanged (approximately 3.5) under various $P_{oxygen}$.

The distribution of O and Ti vacancies is also characterized by electron ptychography (details in methods) which exhibits a linear phase response and outstanding accuracy (Fig. 1f)[24,25]. Phase dips at the O and Ti sites can be visualized as indicated by yellow dashed rectangles in Fig. 1g, h, demonstrating the existence of O and Ti vacancies. Quantitative phase profiles extracted along the dashed arrows within the red and blue boxes of the projected phase images demonstrate significantly weaker O and Ti phase intensities in the red curve (Fig. 1g, h). By statistics on the phase contrast in O and Ti sites from large-area phase image as shown in Supplementary Fig. 5, the direct picture of vacancies distribution has been obtained; we can find that the O vacancies are generally surrounding the Ti vacancies.

Additionally, SHG measurements for bulk STO and the $SrTi_{1-\beta}O_{3-\gamma}$ films (Supplementary Fig. 6) identify the tetragonal phases with a point group of 4$mm$ by theoretical fitting of the rotational anisotropy SHG intensity for all the $SrTi_{1-\beta}O_{3-\gamma}$ films, while the SHG intensity of bulk STO is negligible due to its centrosymmetric cubic lattice structure. Namely, the induction of O and Ti vacancies triggered a crystalline phase transition from cubic to tetragonal, consistent with our XRD and RSM results (Fig. 1a and Supplementary Fig. 1).

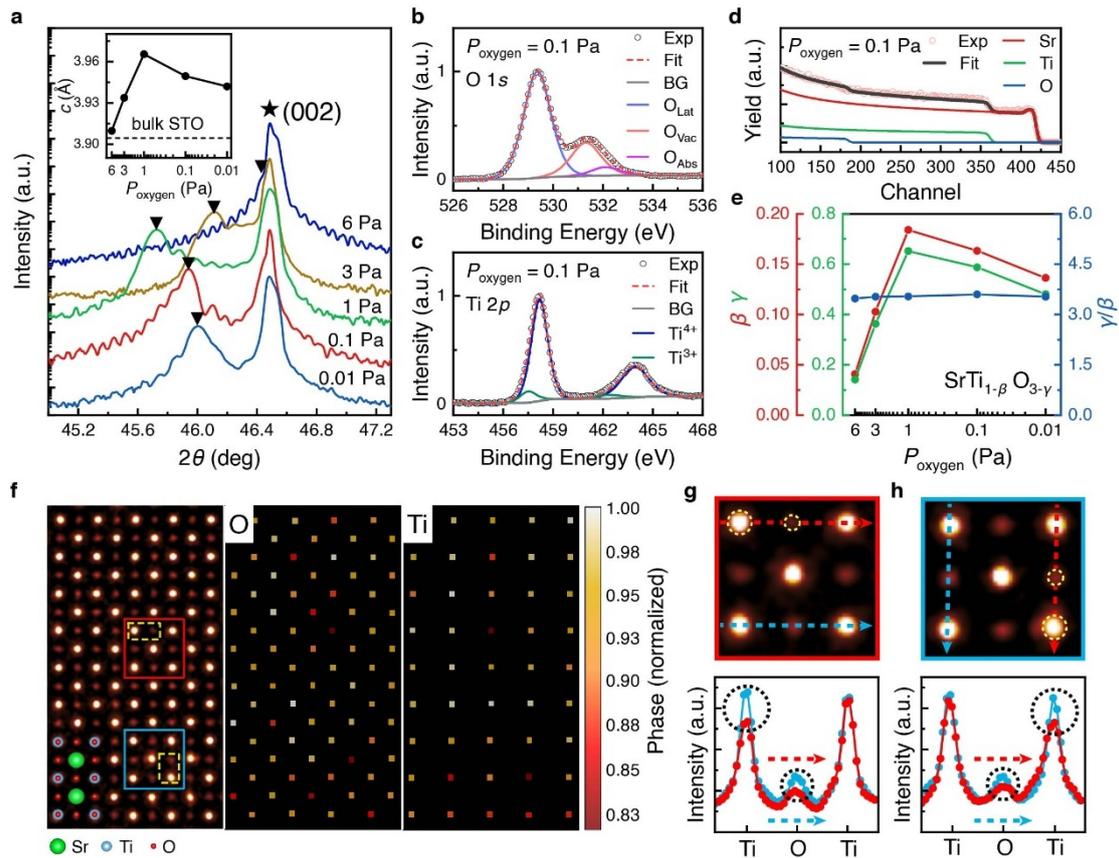

**Fig. 1 | Structure characterization for the $SrTi_{1-\beta}O_{3-\gamma}$ films. a**, X-ray diffraction (XRD) $\theta$-$2\theta$ symmetric scans and (inset) out-of-plane lattice constants $c$ of the $SrTi_{1-\beta}O_{3-\gamma}$ films grown with various oxygen pressures ($P_{oxygen}$). **b, c**, X-ray photoelectron spectroscopy (XPS) fitting results of (**b**) O 1$s$ and (**c**) Ti 2$p$ characteristic spectra for the $SrTi_{1-\beta}O_{3-\gamma}$ films grown with $P_{oxygen}$ = 0.1 Pa (Exp: experiment data; Fit: fitting curves; BG: background signal; $O_{Lat}$: lattice oxygen; $O_{Vac}$: oxygen vacancy; $O_{Abs}$: adsorbed oxygen). **d**, Rutherford backscattering spectroscopy (RBS) fitting result for the $SrTi_{1-\beta}O_{3-\gamma}$ films grown with $P_{oxygen}$ = 0.1 Pa. **e**, Contents of O and Ti vacancies ($\gamma$ and $\beta$), and their ratio ($\gamma/\beta$) in the $SrTi_{1-\beta}O_{3-\gamma}$ films as a function of $P_{oxygen}$. **f**, Phase image of $SrTi_{1-\beta}O_{3-\gamma}$ reconstructed by electron ptychography, in which the neighbored O and Ti vacancies are framed with yellow dashed rectangles. The right two panels are color-coded phase maps for separated O and Ti columns, where red tiny blocks indicate the presence of O and Ti vacancies. **g, h**, Line profiles of phase image from red and blue boxed in (**f**), respectively, with phase dips at the O and Ti sites as highlighted by circles demonstrating the existence of neighboring O and Ti vacancies.

To explore the potential ferroelectricity in the $SrTi_{1-\beta}O_{3-\gamma}$ films, polarization-electric field (*P-E*) hysteresis loop measurements were performed for all the $SrTi_{1-\beta}O_{3-\gamma}$ films at room-temperature (details in methods). As shown in Fig. 2a, the $SrTi_{1-\beta}O_{3-\gamma}$ films with $(\beta, \gamma) = (0.14, 0.48)$, $(0.16, 0.59)$, and $(0.18, 0.65)$ exhibit *P-E* hysteresis loops, demonstrating their ferroelectric nature. The slim hysteresis loops, characterized by small coercive fields and remanent polarization, indicate a relaxor ferroelectricity with weak long-range coupling of polarized domains[26-30]. Figure. 2b presents the *P-E* hysteresis loops for the $SrTi_{1-\beta}O_{3-\gamma}$ films with $(\beta, \gamma) = (0.16, 0.59)$ measured at temperatures ranging from 3 K to 300 K (see Supplementary Fig. 7 for more results), demonstrating the robust ferroelectricity against thermal fluctuation. In addition, the room-temperature maximum polarization ($P_{max}$) and remnant polarization ($P_R$) measured at an electric field of 2.3 MV/cm are $60\ \mu C/cm^2$ and $20\ \mu C/cm^2$, respectively, exceeding the state-of-art values in other STO-based systems (Supplementary Fig. 8)[26,27,31]. Furthermore, the pure hysteresis components at 3 K (Fig. 2c) were also extracted by subtracting the nonhysteresis polarization from the total ones (details in methods)[28,32,33], confirming the relaxor ferroelectricity.

Based on these *P-E* hysteresis loops, a phase diagram for the $SrTi_{1-\beta}O_{3-\gamma}$ films was plotted (Fig. 2d), illustrating $P_{max}$ as functions of temperature and O vacancy content [measured at an electric field of 0.5 MV/cm for the $SrTi_{1-\beta}O_{3-\gamma}$ films with $(\beta, \gamma) = (0.14, 0.48)$, $(0.16, 0.59)$, and $(0.18, 0.65)$, while a reduced electric field of 0.05 MV/cm is adopted for the $SrTi_{1-\beta}O_{3-\gamma}$ films with $(\beta, \gamma) = (0.04, 0.14)$ and $(0.10, 0.36)$ to prevent electrical breakdown under stronger electric fields; see Supplementary Fig. 7 for more results]. The $SrTi_{1-\beta}O_{3-\gamma}$ films with $\gamma \leq 0.36$ exhibit linear dielectric behavior, whereas those with $\gamma \geq 0.48$ display relaxor ferroelectricity. We think the relaxor ferroelectric nature of these $SrTi_{1-\beta}O_{3-\gamma}$ films origins from two aspects. First, the cubic-to-tetragonal crystalline phase transition resulted by O vacancies may enable sizable displacements of Ti atoms[34,35]. Second, the asymmetric distribution of O vacancies surrounding Ti vacancies may generate local dipoles. This local symmetry breaking further stabilize the neighboring dipoles, leading to the formation of nanoscale polar domains. This mechanism was confirmed by atomic-resolved HAADF images (details in methods) on $SrTi_{1-\beta}O_{3-\gamma}$. As shown in Fig. 2e, the off-center displacements of Ti atoms are marked by arrows, showing the nanosized polar domains as a typical feature of relaxor ferroelectricity.

Furthermore, the $SrTi_{1-\beta}O_{3-\gamma}$ films with $(\beta, \gamma) = (0.16, 0.59)$ were selected as examples for the following discussion on the ferroelectric properties (Supplementary Figs. 9-10) owing to their highest $P_{max}$ (Fig. 2a and Supplementary Fig. 8). Piezoresponse force microscopy (PFM) measurements were performed on them at room-temperature, with standard bipolar domain patterns written in a region of 5 μm × 5 μm as framed in Fig. 2f. The well-defined bright and dark domain patterns with nearly 180° phase contrast were observed in the phase image, corresponding to the upward and downward polarization states, respectively. The domain walls were clearly observed at the boundaries between the adjacent polarized regions in the amplitude image (Fig. 2g). Besides, the unwritten region exhibits a virgin domain of upward polarization. Moreover, rotational anisotropy SHG measurements at a set of temperatures (Fig. 2h), as well as temperature-dependent SHG measurements from 300 K to 780 K (Fig. 2i), demonstrate that the SHG intensity gradually decreases with temperature, which is also a typical feature of relaxor ferroelectrics[29], and remains quite strong at 780 K, suggesting the robust relaxor ferroelectricity with a $T_C^{FE}$ higher than 780 K.

Complementing their robust ferroelectric behavior, room-temperature ferromagnetism was also found in all the $SrTi_{1-\beta}O_{3-\gamma}$ films, confirmed by the in-plane magnetic hysteresis loops as shown in Fig. 2j. The saturation magnetization ($M_S$) as a function of O vacancy content is present in the inset of Fig. 2j, suggesting $M_S$ increases with the content of O vacancies and highlighting the crucial role of O vacancies for the emergent ferromagnetism[27,36,37]. Given that $SrTi_{1-\beta}O_{3-\gamma}$ films with $(\beta, \gamma) = (0.16, 0.59)$ exhibit the best ferroelectricity and sufficiently strong ferromagnetism, we still take them as examples for the following discussion on the ferromagnetic properties. The zero-field cooling (ZFC) and field cooling (FC) temperature-dependent magnetization (*M-T*) curves reveal a distinct ferromagnetic-to-paramagnetic transition as shown in Fig.

2k, suggesting robust ferromagnetism with a $T_C^{FM}$ of 830 K, which surpasses previous reported values in other state-of-art STO-based systems[27]. The high $T_C^{FM}$ suggests strong short-range exchange interactions, while the small coercive fields and remanent magnetic moments observed in the magnetic hysteresis loops suggest weak magnetic domain interactions, corroborated by the small initial magnetic domain size (approximately 50 nm) observed in our out-of-plane magnetic force microscope (MFM) results (Fig. 2l). As the applied magnetic field (H) increases, the magnetic domains align with the field direction and eventually become fully magnetized.

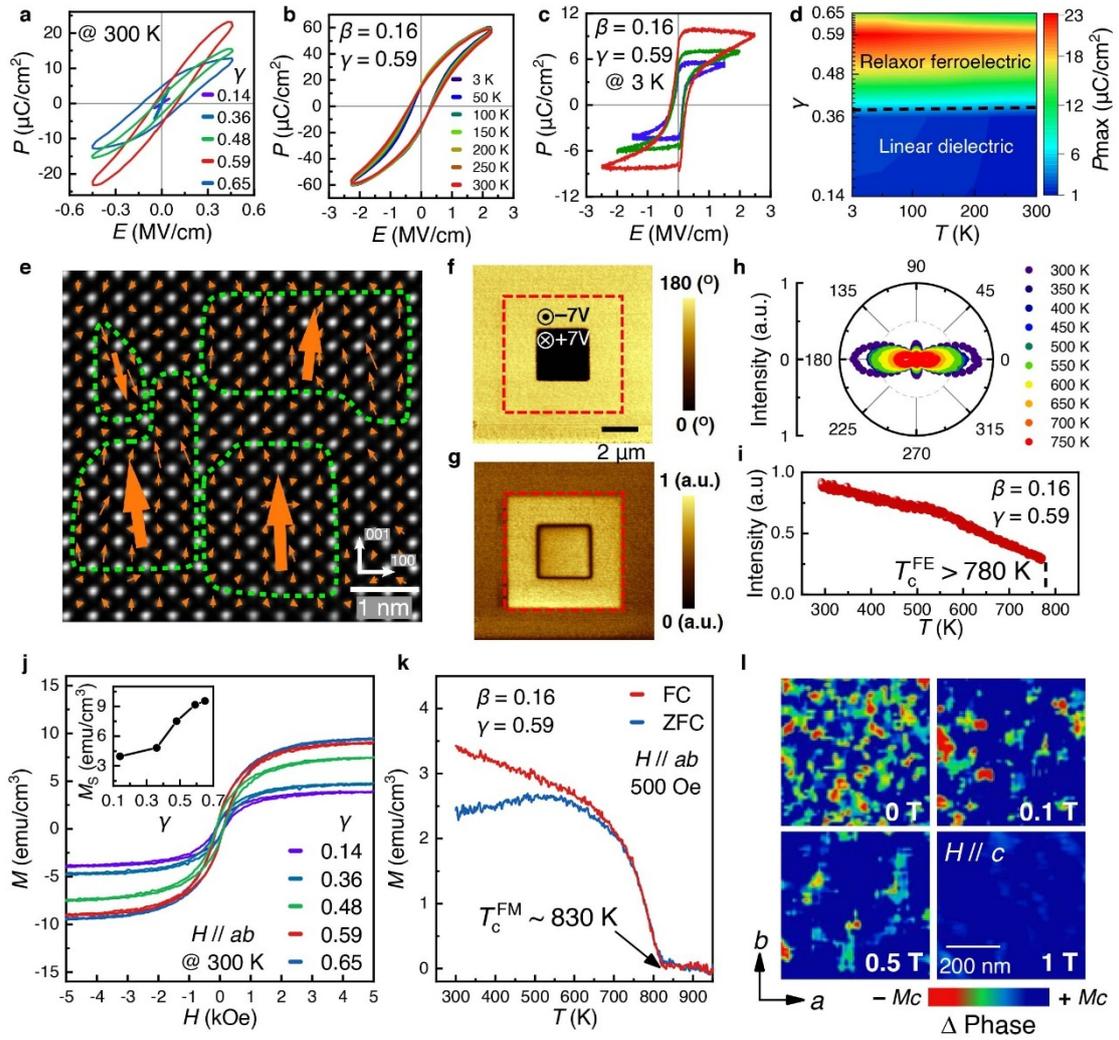

**Fig. 2 | Characterization of ferroelectricity and ferromagnetism for the $SrTi_{1-\beta}O_{3-\gamma}$ films. a**, Polarization-electric field (*P-E*) hysteresis loops of the $SrTi_{1-\beta}O_{3-\gamma}$ films at 300 K. **b**, *P-E* hysteresis loops of the $SrTi_{1-\beta}O_{3-\gamma}$ films with ($\beta$, $\gamma$) = (0.16, 0.59) at various temperatures. **c**, Pure hysteresis

components of the $SrTi_{1-\beta}O_{3-\gamma}$ films with $(\beta, \gamma) = (0.16, 0.59)$ at 3 K. **d**, Ferroelectric phase diagram of the $SrTi_{1-\beta}O_{3-\gamma}$ films, showing the maximum polarization ($P_{max}$) as functions of temperature and O vacancy content [measured at an electric field of 0.5 MV/cm for the $SrTi_{1-\beta}O_{3-\gamma}$ films with $(\beta, \gamma) = (0.14, 0.48)$, $(0.16, 0.59)$, and $(0.18, 0.65)$, while a reduced electric field of 0.05 MV/cm is adopted for the $SrTi_{1-\beta}O_{3-\gamma}$ films with $(\beta, \gamma) = (0.04, 0.14)$ and $(0.10, 0.36)$]. **e**, High-angle annular dark-field (HAADF) image of the $SrTi_{1-\beta}O_{3-\gamma}$ thin film, which is overlayed by the polarization vectors (small arrows) for each unit cell based on the Ti atomic displacements. Nanosized polar domains with similar polarization direction are framed by green dashed lines, the direction of which are indicated by big arrows. **f, g**, Amplitude (**f**) and phase image (**g**) of PFM measurements on the $SrTi_{1-\beta}O_{3-\gamma}$ films with $(\beta, \gamma) = (0.16, 0.59)$, where the standard bipolar domain patterns were written in the framed region. **h**, Rotational anisotropy SHG measurements of the $SrTi_{1-\beta}O_{3-\gamma}$ films with $(\beta, \gamma) = (0.16, 0.59)$ at a set of temperatures ranging from 300 K to 750 K with an interval of 50 K. **i**, SHG measurements of the $SrTi_{1-\beta}O_{3-\gamma}$ films with $(\beta, \gamma) = (0.16, 0.59)$ at continuously varying temperatures ranging from 300 K to 780 K. **j**, Magnetic hysteresis loops of the $SrTi_{1-\beta}O_{3-\gamma}$ films, from which $M_S$ as a function of $\gamma$ are extracted. **k**, Zero-field cooling (ZFC) and field cooling (FC) temperature-dependent magnetization (*M-T*) curves for the $SrTi_{1-\beta}O_{3-\gamma}$ films with $(\beta, \gamma) = (0.16, 0.59)$ as a function of temperature. **l**, Magnetic force microscope (MFM) measurements of the $SrTi_{1-\beta}O_{3-\gamma}$ films with $(\beta, \gamma) = (0.16, 0.59)$ under various magnetic fields of 0 T, 0.1 T, 0.5 T, and 1 T, respectively.

The most significant and engaging findings are the observation of room-temperature ME coupling in the $SrTi_{1-\beta}O_{3-\gamma}$ films. As shown in Fig. 3a, b, magnetic hysteresis loops were measured under varying external electric fields (*E*) at 300 K for the $SrTi_{1-\beta}O_{3-\gamma}$ films with $(\beta, \gamma) = (0.16, 0.59)$ and with thicknesses of 200 nm and 610 nm (Supplementary Fig. 11), respectively (details in methods). $M_S$ of the 200-nm-thick $SrTi_{1-\beta}O_{3-\gamma}$ increases from 9.16 emu/cm$^3$ to 15.75 emu/cm$^3$ as *E* increases from 0 kV/cm to 200 kV/cm, showing strong converse ME coupling. Beyond 200 kV/cm, up to 350 kV/cm, $M_S$ remains unchanged, indicating that the effective range of converse

ME coupling for this thickness is between 0 kV/cm and 200 kV/cm. For the 610-nm-thick films, $M_S$ increases from 12.28 emu/cm³ to 16.24 emu/cm³ as $E$ increases from 0 kV/cm to 82 kV/cm. Notably, the maximum $M_S$ for these $SrTi_{1-\beta}O_{3-\gamma}$ films with different thicknesses are approximately identical, whereas the initial $M_S$ for the thicker one are larger. This discrepancy is attributed to the thickness-dependent initial polarization states, which influence the magnetization and can vary significantly through long-range dipolar interactions. The maximum $M_S$, on the other hand, appears to be primarily determined by the content of O and Ti vacancies.

Under each $H$, magnetization ($M$) varies as a function of $E$. The converse ME coupling coefficient, $\alpha_{CME} = \mu_0 \frac{dM}{dE}$ [38,39], can be obtained as a function of $H$ through linear fitting within the converse ME coupling range as illustrated in Fig. 3c, d. As shown in Fig. 3e, $\alpha_{CME}$ increase with $H$ and approach maximum values of approximately 403 ps/m and 498 ps/m for the 200-nm-thick and the 610-nm-thick films, respectively. Comparing with other single-phase multiferroics such as $Cr_2O_3$ with $\alpha_{CME} = 4 \, ps/m$ obtained at $T \approx 267$ K [7,20] and $TbPO_4$ with $\alpha_{CME} = 37 \, ps/m$ obtained at $T \approx 2$ K [21], we believe that $SrTi_{1-\beta}O_{3-\gamma}$ exhibit significantly stronger converse ME coupling owing to their ferromagnetic (rather than antiferromagnetic) ground state. To be noted, while $BiFeO_3$ has long been considered the only room-temperature single-phase ME multiferroic, its converse ME coupling is rarely reported compared to its direct ME coupling (polarization controlled by magnetic field) perhaps due to its weak magnetization[40-42]. Significantly, ME coupling in $SrTi_{1-\beta}O_{3-\gamma}$ films may persist at an unprecedented-high temperature of 780 K, at which both ferroelectricity and ferromagnetism maintain stability, though direct experimental verification is currently inaccessible due to our instrumental limitations.

In addition, MFM measurements were conducted on the Pt (3 nm)/$SrTi_{1-\beta}O_{3-\gamma}$ (200 nm)/SNTO to capture the evolution of magnetic domains under varying electric fields (details in methods). As shown in Fig. 3f, the initial magnetic state exhibited a multi-domain feature (the larger domain size compared to that in Fig. 2l can be attributed to the enhancement of magnetic domain interactions due to the top electrode, Pt layer). As

*E* increases, the small domains are gradually connected with each other and the domain size eventually exceeds the entire detection area of $2.5\ \mu m \times 2.5\ \mu m$, demonstrating the enhancement of magnetic domain interactions under electric fields.

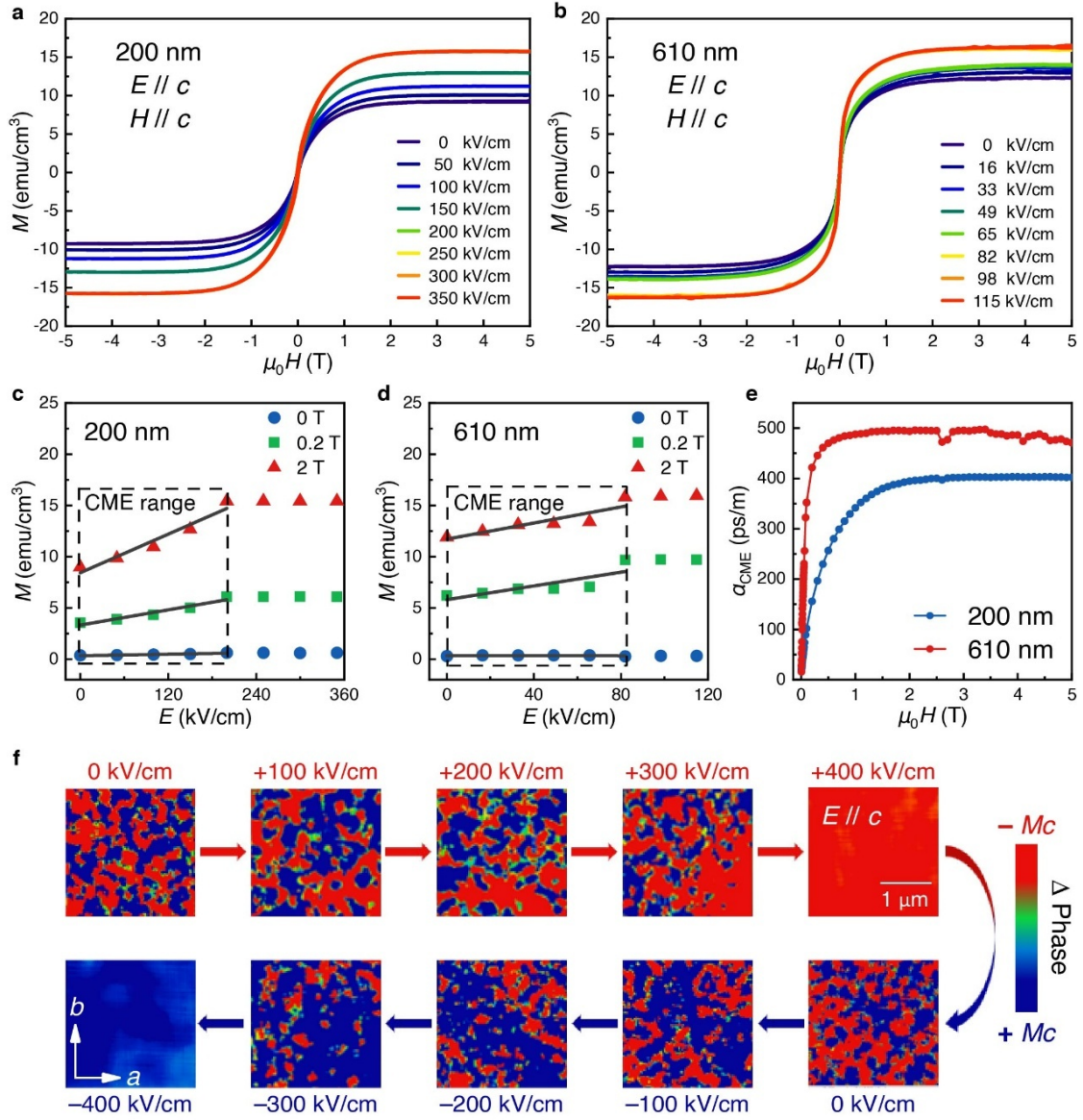

**Fig. 3 | Room-temperature magnetoelectric coupling in the $SrTi_{1-\beta}O_{3-\gamma}$ films with $(\beta, \gamma) = (0.16, 0.59)$. a, b**, Magnetic hysteresis loops of the (**a**) 200-nm-thick and (**b**) 610-nm-thick $SrTi_{1-\beta}O_{3-\gamma}$ films at 300 K under various electric fields. **c, d**, Magnetization (*M*) as a function of *E* for the (**c**) 200-nm-thick and (**d**) 610-nm-thick $SrTi_{1-\beta}O_{3-\gamma}$ films at magnetic fields ($\mu_0 H$) of 0, 0.2, and 2 T, respectively. The converse ME coupling coefficient ($\alpha_{CME}$) are extracted by linear fitting (solid lines) within the converse ME coupling range (CME range). **e**, $\alpha_{CME}$ as a function of $\mu_0 H$. **f**, MFM results of the 200-nm-thick films under varying electric fields.

To reveal the origin of the ferroelectricity, the ferromagnetism, and the microscopic mechanism behind the ME coupling, comprehensive first-principles calculations were performed. First of all, the total energy of a $2\times2\times4$ supercell (s.c.) with one Ti vacancy at the center and one O vacancy at various O-sites, as shown in Fig. 4a, exhibits a sharp drop with the O vacancy locating near the Ti vacancy, giving rise to a potential well around the Ti vacancy that would trap the O vacancy. Namely, O and Ti vacancies in the $SrTi_{1-\beta}O_{3-\gamma}$ films tend to gather, consistent with our electron ptychography results (Fig. 1g, h and Supplementary Fig. 5). We also performed machine-learning accelerated ab-initio molecular dynamic simulations and confirmed the thermal stability for all the structures that contain more than two O vacancies surrounding a Ti vacancy at 300 K (Supplementary Fig. 12).

Then, first-principles calculations were performed on $2\times2\times2$ supercells containing one Ti vacancy each and more than two O vacancies, corresponding to a Ti vacancy content of 0.125 per unit-cell, which falls within the experimental values ($0.04<\beta<0.18$). To descript the location of O vacancies surrounding the Ti vacancy, the O-sites are labeled from 1 to 6, as illustrated in Fig. 4b. The calculations reveal that the structures with asymmetric distributions of O vacancies are energetically favorable, resulting in the spontaneous polarization (Supplementary Table 2). Furthermore, nudged elastic band calculations were performed to address the reversal pathway of polarization, suggesting that O atoms (also O vacancies) tend to hop via neighboring sites rather than directly traverse the Ti vacancies between the diagonal sites. As shown in Fig. 4c, in a $2\times2\times2$ supercell with sites 1 and 6 occupied by O atom, one mechanism to reverse the out-of-plane polarization involves a direct O-atom hopping (1→2), while another proceeds via a two-step process (1→3→2), which exhibits lower energy barrier. On the other hand, the total magnetic moments were calculated to range from 0 $\mu_B$/s.c. to 2 $\mu_B$/s.c. (corresponding to 0~39 emu/cm$^3$), with the experimental values (0~16.24 emu/cm$^3$) falling within this range.

Given the relatively large atomic radius of Ti, the drifting of Ti vacancies can be excluded. Consequently, Ti vacancies act as anchors for O vacancies, preventing them from drifting away. Under external electric fields, O vacancies (also O atoms) hop among the O-sites surrounding Ti vacancies, leading to asymmetric distributions of O vacancies along the electric filed. For instance, a $2\times2\times2$ supercell with sites 3, 4, and 6 occupied by O atoms (symmetric along the electric filed) can transfer to one with sites 2, 3, and 6 occupied (asymmetric along the electric filed) under an out-of-plane electric field. As shown in Fig. 4d, this structural transition results in an electronic configuration rearrangement and an enhancement of magnetic moment ($m$) (Supplementary Table 2).

The symmetric distribution of O vacancies surrounding Ti vacancies are likely to be found in the ferroelectric domain walls, whose proportion can be affected by film thickness which influences the polarization states through long-range dipolar interactions, leading to the thickness-dependent magnetization in $SrTi_{1-\beta}O_{3-\gamma}$ films. As external electric fields increase, the ferroelectric domains align and result in a progressive erasure of domain walls as well as the symmetric distributions, leading to the enhancement of magnetization. Once all the distributions of O vacancies surrounding Ti vacancies transferred from symmetric to asymmetric under electric fields, further increasing the electric field only results in negligible variation of the structure and magnetization, which explains the plateau of $M$ beyond the converse ME coupling range in Fig. 3c,d.

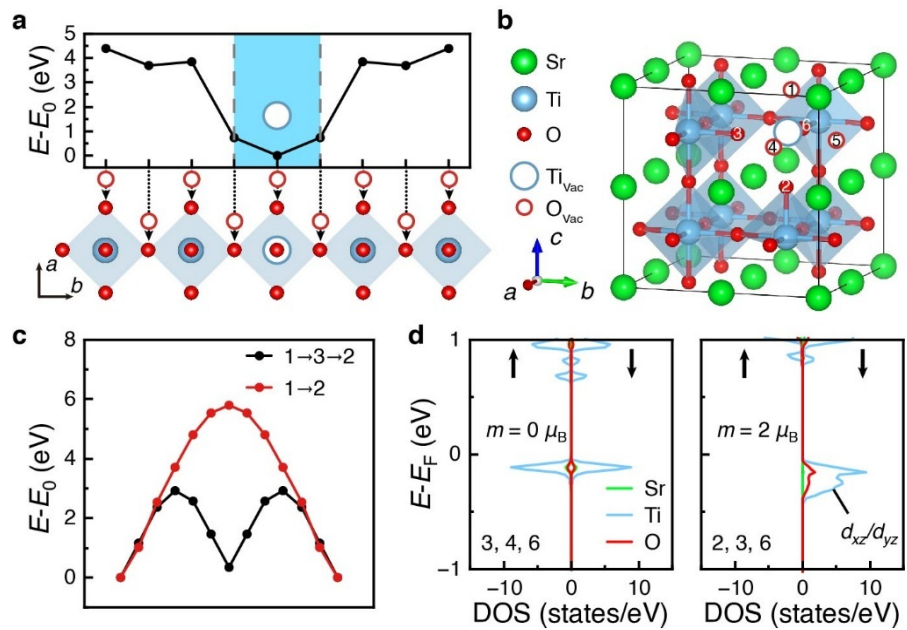

**Fig. 4 | First-principles calculation results. a**, Total energy ($E$-$E_0$) of a $2\times2\times4$ $SrTi_{1-\beta}O_{3-\gamma}$ supercell with a Ti vacancy ($Ti_{Vac}$) at the center and an O vacancy ($O_{Vac}$) at various O-sites. **b**, $2\times2\times2$ $SrTi_{1-\beta}O_{3-\gamma}$ supercell with O vacancies surrounding a Ti vacancy where O-sites surrounding the Ti vacancy are labeled from 1 to 6. **c**, Nudged elastic band of possible pathway for out-of-plane polarization reversal in a $2\times2\times2$ $SrTi_{1-\beta}O_{3-\gamma}$ supercell containing a Ti vacancy, around which the O-

sites 1 and 6 (2 and 6) are occupied corresponding to the initial (final) polarization state. **d**, Densities of states (DOS) for a $2 \times 2 \times 2$ $SrTi_{1-\beta}O_{3-\gamma}$ supercells with O vacancies surrounding a Ti vacancy, where the O atoms occupy the sites 3, 4, and 6 (left) and sites 2, 3, and 6 (right), respectively.

We have demonstrated the coexistence of ferroelectricity and ferromagnetism in the $SrTi_{1-\beta}O_{3-\gamma}$ films by dedicated targeted vacancy engineering, with $T_C^{FE}$ higher than 780 K and $T_C^{FM}$ of 830 K. More importantly, strong converse ME coupling was achieved with $\alpha_{CME} = 498\,p\,m/s$ at room-temperature, which is superior to the previously reported single-phase ME multiferroics. A microscopic mechanism for the ME coupling was proposed by our atomic-scale structural analysis and first-principles calculations, where the ferroelectricity is triggered by a cubic-to-tetragonal crystalline phase transition and the asymmetric distribution of O vacancies surrounding the Ti vacancies, and the ferromagnetism originates from charge injection induced by O vacancies. Furthermore, the hopping of O vacancies under external electric field results in an electronic configuration rearrangement, leading to an enhancement of magnetization. These findings open new avenues for discovering high-performance single-phase ME multiferroics within the vast family of perovskite oxides.

## Methods

### Target Synthesis and Film Growth

All the films presented in this work were epitaxially grown on the $(001)$-$Sr_{1-x}Nb_xTiO_3$ ($x = 0.007$) substrates at 780 °C via pulsed laser deposition with various oxygen pressure (6 Pa, 3 Pa, 1 Pa, 0.1 Pa, and 0.01 Pa, respectively). The stoichiometric SrTiO$_3$ target was synthesized by sintering the uniform mixture of SrCO$_3$ and TiO$_2$ powder with a precise molar ratio 1:1 for 15 h at 1250 °C, 1300 °C, and 1350 °C, successively. The laser with a central wavelength of 308 nm, a power of 3 J/cm$^2$, and a repetition frequency of 4 Hz, was provided by a XeCl excimer laser. After growth, the films were cooled down to room-temperature under the oxygen pressure with a rate of 10 °C/min.

### Crystal Structure Characterization

The X-ray diffraction (XRD) $\theta$-$2\theta$ symmetric scans, reciprocal space mapping (RSM), and X-ray reflectivity (XRR) measurements were carried out using a PANalytical X'Pert3 MRD diffractometer, with the laser wavelength of 0.154 nm. All XRR results were fitted using GenX software.

### Elemental Stoichiometric Analysis

Elemental stoichiometric analysis of the bulk STO and the $SrTi_{1-\beta}O_{3-\gamma}$ films grown under varying oxygen pressures were performed using Rutherford backscattering spectrometry (RBS). The RBS experiments were conducted at the Heavy Ion Institute of Peking University, utilizing 2 MeV helium ions incident on the samples with a scattering angle of 160°, an exit angle of 20°, and an energy resolution of 20 keV. The elemental proportions for each sample were extracted by accurately fitting the RBS spectra with the Simnra software.

### Atomic-scale Structural Characterization

The atomic-scale structural characterization of the thin film was conducted using a spherical aberration-corrected transmission electron microscope JEM NeoARM200 (JEOL Ltd., Tokyo) operated at an accelerating voltage of 200 kV. The electron ptychography data were acquired on a high-speed direct electron detection camera

(DECTRIS ARINA) which enables high-quality data acquisition at speeds comparable to conventional scanning transmission electron microscope (STEM) measurements. The iterative reconstruction of electron ptychography datasets was performed using the foldslice algorithm *(24)*. High-angle annular dark-field (HAADF) image was obtained using a collection angles of 75-310 mrad.

**Piezoresponse Force Microscope (PFM) Measurements**

PFM measurements were performed on $SrTi_{1-\beta}O_{3-\gamma}$ using a scanning probe microscope (Asylum Research MFP-3D). PFM imaging was carried out with Nanosensor PPP-EFM-10 Ir/Pt-coated conductive tips. Local hysteresis behaviors of PFM phase and amplitude signals were recorded in DART mode (dual AC resonance tracking) with the signals collected in the absence of applied voltage. Domain writing was performed with a switching voltage of ±7 V.

**Polarization-electric Field (*P-E*) Hysteresis Loop Measurements**

A home-made sample rod was used to connect the precision multiferroic analyzer (RADIANT Tec. Inc.) with the physical property measurement system (PPMS), enabling temperature-dependent *P-E* hysteresis loop measurements (3~300 K) for the 200-nm-thick $SrTi_{1-\beta}O_{3-\gamma}$ films grown under various oxygen pressures. Round Pt electrodes (with 50 nm thick and 30 μm diameter) were deposited using a magnetron sputtering system and patterned with standard microfabrication techniques. The *P-E* hysteresis loops were obtained in a simple hysteresis mode, where triangular pulses with a frequency of 10 kHz were applied to reverse the polarization of the $SrTi_{1-\beta}O_{3-\gamma}$ layer. The PUND (positive-up-negative-down) measurements were conducted with the standard mode of the precision multiferroic analyzer (RADIANT Tec. Inc.).

**Second Harmonic Generation (SHG) Measurements**

Optical SHG measurements were conducted on the bulk STO and the $SrTi_{1-\beta}O_{3-\gamma}$ films, respectively. The SHG signals were obtained in reflection geometry. An 800-nm-wavelength laser with a power of 80 mW incident at an angle of 45° relative to the surface normal was provided by a Ti:Sapphire femtosecond laser (Tsunami 3941-X1BB, Spectra-Physics) and served as the pump beam. The polarization direction ($\phi$) of the incident light was controlled via an automated $\lambda/2$ wave plate. The second harmonic fields ($E_{2\omega}$) generated within the films through nonlinear optical processes were decomposed into p- ($I^{2\omega}_{p-\text{out}}$) and s- ($I^{2\omega}_{s-\text{out}}$) components by a polarizing beam splitter. The resulting optical signals were detected by a photomultiplier tube. The SHG polarimetry data were analyzed and fitted using theoretical models based on the standard point group (4*mm*) symmetry.

**X-ray Photoelectron Spectroscopy (XPS) Measurements**

XPS measurements were conducted using an X-ray photoelectron spectrometer (ESCALAB 250Xi, Thermo Fisher Scientific). All elemental characteristic spectra were calibrated to the C 1*s* binding energy of 284.8 eV and subsequently fitted with the CasaXPS software.

**Macroscopic Magnetic Properties Measurements**

The macroscopic magnetic properties were measured on a Quantum Design vibrating sample magnetometer (VSM-SQUID). All magnetic hysteresis loop measurements were performed at 300 K. The magnetization-temperature curves of $\text{SrTi}_{1-\beta}\text{O}_{3-\gamma}$ were performed with a high-temperature module in a temperature range of 300~950 K. Both zero-field cooling (ZFC) and field cooling (FC, 500 Oe) M-T curves were recorded at a rate of 10 K/min. A layer of Pt with thickness of 50 nm was deposited on the $\text{SrTi}_{1-\beta}\text{O}_{3-\gamma}$ as a top electrode via magnetron sputtering to enable the application of electric fields. After each measurement, the sample was demagnetized under the applied voltage. Subsequently, the voltage was adjusted for the next measurement, and the process was repeated.

**Microscopic Magnetoelectric Coupling Measurements**

Magnetic force microscope (MFM, attoDRY2100) was utilized to investigate the evolution of microscopic magnetic domains of $\text{SrTi}_{1-\beta}\text{O}_{3-\gamma}$ under progressively applied electric and magnetic fields at room-temperature. A layer of Pt with thickness of 3 nm was deposited on the $\text{SrTi}_{1-\beta}\text{O}_{3-\gamma}$ as a top electrode via magnetron sputtering to enable the application of electric fields. All the measurements were conducted in vacuum based on a phase modulation technique in noncontact AC mode, with a cantilever with a spring constant *k* of 2.8 N/m and a resonant frequency *f* of 75 kHz. In the AC mode, the cantilever is mechanically excited at its natural resonance frequency with an initial phase. The attractive/repulsive magnetic interaction between the cantilever and sample surface will cause a negative/positive phase shift. Thus, the phase mapping image can give the local information on the *z*-direction magnetization of a ferromagnetic film.

**First-principles Calculation**

All first-principles calculations are based on density functional theory (DFT) as implemented in the Vienna ab initio simulation package[43,44], using the projector augmented-wave method[45]. The exchange-correlation potential is adopted by the generalized gradient approximation of the Perdew-Burke-Ernzerhof functional for solids[46]. The DFT+U method is adopted to improve the description of on-site Columb interactions of the Ti-3*d* orbitals with $U_{\text{eff}}$ set to 4 eV[47]. The plane-wave cutoff energy is set to 520 eV. Γ-centered k-meshes of $4\times4\times4$, and $8\times8\times2$ are adopted for $2\times2\times2$ and $1\times1\times4$ supercells, respectively. The out-of-plane lattice parameters and the ionic positions are fully optimized until the forces on the ions are less than $10^{-2}$ eV/Å, while the in-plane lattice constants are fixed as 3.905 Å. The energy convergence criterion is set to $10^{-6}$ eV. The machine-learning accelerated ab-initio molecular dynamic simulations are performed on $4\times4\times4$ supercells with $2\times2\times2$ O-Ti cluster sublattices for 20000 steps with a time step of 0.5 fs. In such a large system (more than 300 atoms), only Γ point is considered to accelerate the calculation. The temperature is set to 300 K in a canonical (NVT) ensemble with a Nose-Hoover thermostat.

## Data availability

Source data of the main text are provided with this paper. Any additional data that support the findings of this study are available from the corresponding author upon reasonable request.

## Competing interests

The authors declare no competing interests.

# Method references

**Acknowledgments** This work is supported by the National Key Research and Development Program of China (Grant No. 2024YFA1409500), the National Key Basic Research Program of China (Grant No. 2020YFA0309100, and 2021YFA1400701), the National Natural Science Foundation of China (Grant No. 11974390, 52322212, 12174437, 12222414, and 12074416), the Youth Innovation Promotion Association of the Chinese Academy of Sciences (Grant No. Y2022003).

**Author contributions** K.-J.J. conceived of the idea and supervised the project; Z.Y., C.M., and M.W. contributed equally to this work; Z.Y., C.M., Q.Z., and K.-J.J. prepared the manuscript; Z.Y. prepared the epitaxial films, performed the crystal structure characterizations, and analyzed the XPS and RBS results, with the supervision of K.-J.J.; Y.H. and Q.Z. provided the atomic-scale structure characterizations; S.X. conducted the second harmonic generation measurements with the supervision of K.-J.J.; Z.Y. and Y.Y. performed the P-E hysteresis loops and PFM measurements with the supervision of K.-J.J.; Z.Y. and M.W. performed the M-H hysteresis loops measurements; M.W. conducted the MFM measurements with the supervision of Y.C. and F.H.; C.M. performed the first-principles calculations and developed the theoretical frameworks, with the supervision of K.-J. J.; C.C., E.-J.G., C.W., X.X., and G.Y. participated in the discussion; all authors were involved in the data analysis and manuscript preparation.